%% file: nautilus-compass.tex
\documentclass{article}

\usepackage[a4paper, margin=1in]{geometry}
\usepackage[utf8]{inputenc}
\usepackage[T1]{fontenc}
\usepackage{lmodern}  
\usepackage{microtype}
\usepackage{graphicx}
\usepackage{booktabs}
\usepackage{multirow}
\usepackage{hyperref}
\usepackage{xcolor}
\usepackage{listings}
\usepackage{url}
\usepackage{caption}
\usepackage[round]{natbib}
\usepackage{amsmath}

\hypersetup{colorlinks=true, linkcolor=blue, urlcolor=blue, citecolor=blue}

\title{Nautilus Compass:\\
       Black-box Persona Drift Detection for Production LLM Agents}

\author{Chunxiao Wang \\
        Yiluo Technology Co., Ltd. \\
        \url{github.com/chunxiaoxx/nautilus-compass} \\
        \texttt{chunxiaoxx@gmail.com}}

\date{\today}

\begin{document}
\maketitle

\begin{abstract}
\input{sections/00_abstract}
\end{abstract}

\section{Introduction}
\input{sections/01_intro}

\section{Related Work}
\input{sections/02_related}

\section{Method}
\input{sections/03_method}

\section{Evaluation}
\input{sections/04_eval}

\section{Discussion}
\input{sections/05_discussion}

\section{Limitations and Future Work}
\input{sections/06_limitations}

\section{Open Source and Reproducibility}
\input{sections/07_opensource}

\bibliographystyle{plainnat}
\bibliography{refs}

\end{document}

%% file: sections/00_abstract.tex

Production LLM coding agents drift over long sessions: they forget
user-specified constraints, slip into mistakes the user already
flagged, and confabulate prior agreements. White-box approaches such
as persona vectors require model weights and so cannot be applied to
closed APIs (Claude, GPT-4) that most users actually interact with.

We present \textbf{Nautilus Compass}, a black-box persona drift
detector and agent memory layer for production coding agents. The
method operates entirely at the prompt-text layer: cosine similarity
between user prompts and behavioral anchor texts, aggregated by a weighted top-$k$ mean using
BGE-m3 embeddings. Compass is, to our knowledge, the only public
agent memory layer (among Mem0, Letta, Cognee, Zep, MemOS, smrti
verified May~2026) that does not call an LLM at index time to extract
facts or build a graph; raw conversation text is embedded directly.
The system ships as a Claude Code plugin, an MCP~2024-11-05 A2A server
(Cursor, Cline, Hermes), a CLI, and a REST API on one daemon, with a
Merkle-chained audit log for tamper-evident anchor updates.

On a held-out test set built from real Claude Code session traces and
labeled by an independent LLM judge, Compass reaches ROC AUC 0.83 for
drift detection. The embedded retrieval pipeline scores 56.6\% on
LongMemEval-S v0.8 and 44.4\% on EverMemBench-Dynamic (n=500), topping
the four published EverMemBench Table~4 baselines. LongMemEval-S 56.6\% is \textasciitilde 30 points below recent
white-box leaders (90+\%); we treat that as the architectural ceiling
of the no-extraction design. End-to-end reproduction cost is \$3.50
(\textasciitilde 14$\times$ cheaper than GPT-4o-judged stacks). A paired cross-vendor behavior A/B
accompanies these numbers as preliminary system-level evidence.

Code, anchors, frozen test data, and audit-log tooling are
MIT-licensed at \url{github.com/chunxiaoxx/nautilus-compass}.

%% file: sections/01_intro.tex

Modern coding agents based on Large Language Models (LLMs) such as
Claude Code, Cursor, and Continue.dev sustain dialogue sessions
spanning hundreds to thousands of turns. Within such sessions, a
recurring failure mode is \emph{persona drift}: the assistant
gradually forgets user-specified constraints (``always verify before
claiming success''), slips into bad habits the user explicitly
flagged earlier (``don't fabricate prior agreements''), or confabulates
having reached agreements that were never made. These behaviors are
not failures of factual recall---retrieval-augmented memory plugins
\citep{mem0,letta} successfully retrieve the relevant memos---but
failures of \emph{behavioral consistency}: the assistant has the
information but does not act on it.

\paragraph{Existing approaches fall into two categories.}
\textit{(a) Memory plugins} like mem0 \citep{mem0}, Letta
\citep{letta}, and Zep \citep{zep} focus on retrieval quality:
finding the right memory entries to inject into the prompt. While
these systems produce strong retrieval metrics (e.g., on
LongMemEval \citep{longmemeval2024}), they leave open the question
of whether the LLM \emph{honors} the retrieved content.
\textit{(b) White-box safety methods}, exemplified by
Persona Vectors \citep{personavectors2025}, identify directions in
the model's activation space corresponding to specific traits and
enable monitoring or steering of trait shifts. However, these
methods require access to model weights or hidden states, putting
them out of reach of the typical end-user who interacts with
proprietary LLMs through black-box APIs.

\paragraph{Our contribution.} We present \textbf{Nautilus Compass}, a
black-box persona drift detection system that runs entirely in user-space
hooks of an LLM coding agent. The reference implementation ships as a
Claude~Code plugin and as a generic MCP~2024-11-05 server (compatible with
Cursor, Cline, Hermes, OpenClaw, and any other MCP client), with a CLI
and a REST endpoint backed by the same BGE-m3 daemon---a single
embedder service amortized across every entry point. The system is the
first product in the \emph{Nautilus} open agent platform, designed so
that any coding agent runtime can obtain drift telemetry without weight
access. At each user-issued prompt, we compute the cosine similarity
between the prompt and a small set of \emph{behavioral anchor texts},
aggregated via weighted top-$k$ mean. Anchors come in two flavors:
positive (desired task patterns) and negative (mistakes the user has
flagged). The difference between aggregated positive and negative
similarity yields a continuous \emph{drift score}, which we threshold
into a three-band output (\emph{aligned} / \emph{neutral} /
\emph{deviation}) and inject into the LLM's context for the current
turn.

\begin{figure}[h]
\centering
\includegraphics[width=0.95\textwidth]{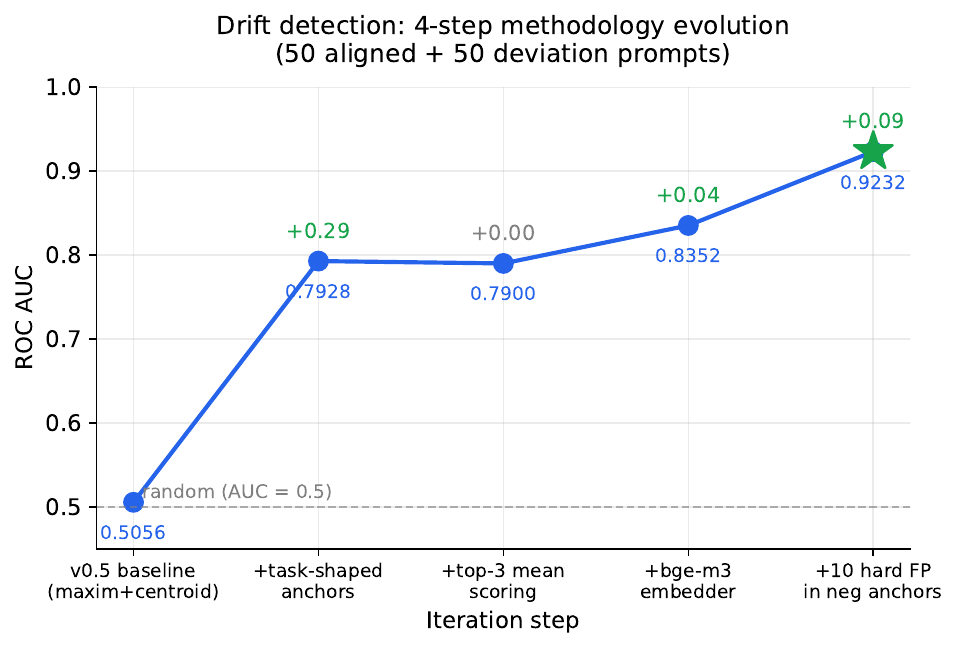}
\caption{Drift detection ROC AUC across the four-step methodology
evolution. Each step is a controlled change with all other
variables fixed; aggregate improvement is from random ($0.5056$)
to strong ($0.9232$). The largest single jump comes from rewriting
abstract maxim anchors as task-shaped sentences matching the
prompt distribution.}
\label{fig:auc-evolution}
\end{figure}

\paragraph{Methodological insight.}
Our central finding is that the design of the anchor set, not the
choice of embedder or aggregation function, is the dominant factor
determining detection accuracy. We document a four-step iteration
that improves ROC AUC from 0.51 (random) to 0.92 on a 100-prompt
synthetic test set (Figure~\ref{fig:auc-evolution}), and report
negative results (e.g., that substituting a cross-encoder reranker
for the bi-encoder cosine yields no measurable improvement at
$36\times$ the latency).

\paragraph{Open source and reproducibility.}
All code, anchor sets, evaluation scripts, and benchmark data are
released under MIT (anchors under CC0) at
\url{https://github.com/chunxiaoxx/nautilus-compass}. The four-step
ablation, the head-to-head comparison with mem0 on LongMemEval-S,
and the negative cross-encoder result are all reproducible from
the repository in under one hour on a CPU-only machine.

\paragraph{Relation to Persona Vectors.}
Our work is \emph{complementary, not competing}, with
\citet{personavectors2025}. Persona Vectors operates in
activation space (white-box), giving more fine-grained signal but
requiring model weights. nautilus-compass operates in prompt-text space
(black-box), giving coarser signal but deployable by anyone with
hook-level access to the agent. We argue that a robust safety
posture for production LLM agents will combine both layers.

\paragraph{Scope and limits of claims.}
This paper makes a claim about \emph{detection accuracy}: given a
user-issued prompt and a curated anchor set, our method classifies
whether the prompt aligns with task-shaped positive patterns or
deviates toward known failure modes, with held-out AUC reported in
Section~\ref{sec:eval-holdout}. We additionally report a
\emph{cross-vendor behavior-steering A/B}
(Section~\ref{sec:limitations-behavior}) across six production LLMs
(Google Gemini-2.5 \{pro, flash\}, MiniMax M2.7,
ByteDance doubao-seed-2.0-pro, DeepSeek v3.2, Zhipu GLM-5.1) judged
by an independent seventh (Moonshot Kimi-k2.6). The cross-vendor
result is mixed but informative: drift injection produces a
statistically significant improvement on fabrication-resistance
($p < 0.05$, $n=120$) and no significant aggregate effect on the
other three axes; one axis (refuse-destructive) trends nominally
negative, which we attribute to the alert text verbalizing the
matched negative anchor. Readers expecting
``nautilus-compass causally improves Claude's safety'' should treat
this paper as preliminary evidence with axis-specific effects, not as
a uniform guarantee.

%% file: sections/02_related.tex

\subsection{Memory Systems for LLM Agents}
\label{sec:related-memory}

A growing line of work addresses the limited context window of LLMs
through external memory. \textbf{mem0} \citep{mem0} stores
key-value memories extracted from conversations and retrieves them
via dense embeddings; the system targets production agent use cases
and reports retrieval Recall@5 in the 0.5--0.6 range on
LongMemEval-S. \textbf{Letta} (formerly MemGPT) \citep{letta}
introduces a hierarchical memory architecture with archival
storage and core memory, modeling LLM context management as an
operating-system problem. \textbf{Zep} \citep{zep} adds temporal
knowledge-graph structure on top of dense retrieval, optimizing for
multi-session temporal reasoning. \textbf{A-MEM} \citep{amem}
introduces dynamic links between memory entries with supersede
detection, addressing memory staleness.

These systems share a common assumption: \emph{the LLM will honor
retrieved memory once it is injected into the context}. We argue
that this assumption fails systematically over long sessions, and
that complementary mechanisms are needed to monitor whether the
agent's behavior actually aligns with the retrieved guidance.

\subsection{White-box Persona Monitoring}
\label{sec:related-persona}

\citet{personavectors2025} represents the state of the art in
white-box persona monitoring: they identify linear directions in
the activation space of an LLM corresponding to traits such as
sycophancy, hallucination propensity, and harmful intent. By
projecting hidden activations onto these directions during
inference, the authors demonstrate the ability to monitor and
steer trait expression at deployment time, predict trait shifts
induced by finetuning, and flag training data likely to cause
undesirable persona changes.

The chief limitation, from the perspective of the typical
end-user of a coding agent, is that this method requires access
to model weights or hidden states. This puts it out of reach of
users interacting with proprietary LLMs (Claude, GPT-4, Gemini)
through API or product surfaces. Our work targets exactly this
deployment gap: providing a black-box analog that runs in
user-space hooks without model-internal access.

\subsection{Black-box Safety and Behavior Classifiers}
\label{sec:related-blackbox}

To our knowledge, no published work targets prompt-level
\emph{persona drift} detection in the closed-API setting that
Persona Vectors addresses with weight access. Adjacent black-box
threads we surveyed:

\textbf{Constitutional AI} \citep{constitutional2022} trains models
to self-critique against natural-language principles. The
principles act as soft anchors at training time, whereas our
anchors operate at inference time over arbitrary deployed models.
\textbf{Detoxify} and similar text classifiers \citep{detoxify}
provide black-box safety classification (toxicity, severe
toxicity, identity attacks) by training discriminative classifiers
on labeled corpora. These give a single safety score per prompt
but do not offer the per-anchor provenance needed to surface a
specific drift mode (e.g., ``you are repeating mistake X you
previously flagged'').
\textbf{OpenAI Moderation API} \citep{openai-mod} and
\textbf{Anthropic Safe API} \citep{anthropic-safe} expose
black-box safety endpoints, but their dimensions
(harassment, self-harm, sexual, violence) are coarser than
production agents' actual drift modes (forgetting verification,
fabricating prior agreement, ignoring user-flagged constraints).
\textbf{Prompt-level safety probes} \citep{promptinject2023,
textdefendr2024} test models against adversarial prompt
injections; this targets a different threat model (external
adversary) from ours (internal session drift).
\textbf{Reward models} as black-box judges
\citep{skywork2024,helpsteer2024} train scalar safety/helpfulness
predictors; these offer one global score per prompt but again
lack per-pattern provenance.

Compared to the above, our contribution is a
\emph{user-space, anchor-based, per-prompt drift detector with
matched-anchor provenance}. The anchors can be authored by
end-users (rather than baked in at training time), making the
system extensible to drift modes the safety vendors do not
prioritize.

\subsection{Embedders and Cross-Encoder Rerankers}
\label{sec:related-embed}

We build on the BGE family of embedding models \citep{bgem3}.
BGE-m3 in particular offers multilingual support across 100+
languages with a 1024-dimensional output, making it suitable for
mixed Chinese/English production agent contexts. For retrieval
reranking, we leverage \texttt{bge-reranker-v2-m3}
\citep{bgereranker}, a cross-encoder trained on similar data,
which scores (query, candidate) pairs jointly rather than
embedding them independently. Foundational sentence embedding
methodology follows \citet{sbert}.

A natural question is whether a cross-encoder can substitute for
the bi-encoder cosine in our drift detection step (anchor
matching). In Section~\ref{sec:eval-rerank-null}, we report a
\emph{negative} result: cross-encoder substitution improves drift
AUC by only 0.0008 at 36$\times$ the latency, suggesting that the
short, semantically heterogeneous nature of behavioral anchor
text leaves little token-interaction signal for cross-encoders to
exploit, in contrast to the long-document setting of retrieval
reranking.

\subsection{Long-Context Memory Benchmarks}
\label{sec:related-bench}

We evaluate retrieval performance using LongMemEval-S
\citep{longmemeval2024}, which provides 500 question-haystack
pairs across six question types: knowledge update, multi-session
synthesis, single-session-user/assistant/preference factual
recall, and temporal reasoning. Each haystack contains roughly 50
candidate sessions, making top-5 retrieval a meaningful target
metric. Other relevant benchmarks include LoCoMo (long
conversation memory) and PerLTQA (Chinese long-term persona
dialogue), which we leave for future evaluation.

\subsection{Anchor-Based Behavior Detection (Analogies)}
\label{sec:related-anchors}

Our anchor-based approach draws conceptual lineage from several
threads. \textbf{Reward modeling in RLHF} aggregates many
preference comparisons into a scalar reward; in our case we
aggregate cosine similarity to many positive/negative behavioral
anchors into a scalar drift score. \textbf{Behavioral cloning}
in robotics learns from demonstrations; our positive anchors
function as ``demonstrations of correct task patterns''.
\textbf{Few-shot in-context learning} relies on a small set of
examples to bias the LLM's distribution; our negative anchors
function as ``examples of behaviors to avoid'', albeit operating
externally rather than in-context.

The core innovation in our system is not any single technical
component but the empirical methodology for designing anchor sets
that yield strong detection signal (Section~\ref{sec:method-evolution}),
together with the integration with hook-level production
deployment.

\subsection{Strategy Distillation}
\label{sec:related-strategy}

\citet{dptagent} introduces distillation path-triggered (DPT)
agents, which extract reasoning patterns from past task
completions and inject them when similar tasks recur. We
incorporate a lightweight DPT-style mechanism for cross-session
strategy retrieval, though it is not the focus of this paper. Our drift detection is
orthogonal to and compatible with strategy distillation.

%% file: sections/03_method.tex

\subsection{Architecture Overview}
\label{sec:method-arch}

nautilus-compass operates as a plugin to Claude Code via three hook
points exposed by the agent runtime: \texttt{UserPromptSubmit}
(fires before each user-issued prompt is processed),
\texttt{PostToolUse} (fires after each tool invocation), and
\texttt{Stop} (fires at session end). Hooks return text payloads
that are injected into the agent's system prompt for the
subsequent turn, with no modification to the underlying LLM.

\begin{figure}[h]
\centering
\includegraphics[width=0.95\textwidth]{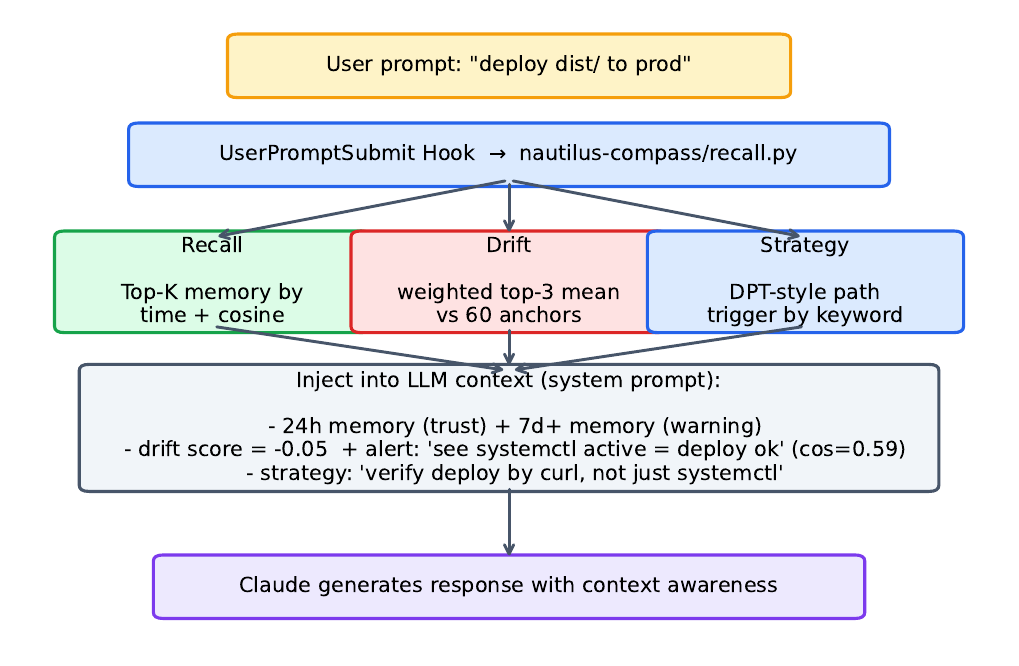}
\caption{System architecture. A user-issued prompt enters
\texttt{UserPromptSubmit}, which forks three parallel paths:
top-$k$ memory recall, drift scoring against behavioral anchors,
and DPT-style strategy retrieval. The combined output is injected
back into the LLM's system prompt for the same turn.}
\label{fig:architecture}
\end{figure}

A persistent BGE embedder daemon \citep{bgem3} runs on a local
TCP socket (\texttt{127.0.0.1:9876}), keeping the embedder warm
across hook invocations. Cold-start latency on a CPU-only
Windows/Linux host is 12--30 seconds; warm hot-path latency is
1.8 seconds (single forward pass plus anchor cache lookup).
Figure~\ref{fig:architecture} illustrates the data flow.

The system computes three artifacts at each
\texttt{UserPromptSubmit}, in parallel where possible:
\textbf{(1)} top-$k$ memory recall against the project's memory
file directory, \textbf{(2)} drift score against behavioral
anchors, and \textbf{(3)} relevant strategy retrieval. The three
are concatenated and injected as a single recall block.

\subsection{Anchor Schema}
\label{sec:method-anchors}

A behavioral anchor is a short natural-language string
(typically 8--40 tokens) describing either a desired behavior
pattern (\emph{positive}) or a flagged failure mode
(\emph{negative}). We adopt a JSON schema with backwards
compatibility:

\begin{verbatim}
{
  "positive_anchors": [
    "I first grep memory to verify whether this issue
     has been discussed before",
    {"text": "I check the deploy by curl-ing the live
     URL, not just systemctl status",
     "weight": 1.0, "tp": 0, "fp": 0},
    ...
  ],
  "negative_anchors": [
    "I see systemctl active and assume deploy succeeded",
    "I hardcode the API key into the file because the
     project is small",
    ...
  ]
}
\end{verbatim}

Each anchor has an associated weight (default 1.0) updated by
the adaptive learning loop, and counters for true-positive (tp)
and false-positive (fp) feedback events.

\subsection{Design Principle: Task-shaped Authoring}
\label{sec:method-task-shape}

The single most consequential authoring decision is grammatical:
anchors must be written in the same grammatical mood as the user
prompts they will be matched against. We call anchors that satisfy
this constraint \emph{task-shaped}.

Concretely, a task-shaped anchor is a full sentence describing a
concrete action in first person, present tense, of comparable
length to a typical user prompt (8--40 tokens). Declarative
maxims (``simplicity over cleverness'', ``verify before claiming
done'') are \emph{not} task-shaped: their abstract grammatical
form does not align with the question/imperative form of typical
user prompts, and bi-encoder cosine similarity collapses across
both aligned and deviation prompts.

We discovered this principle empirically (Section~\ref{sec:method-evolution}
documents the ablation): rewriting an entire anchor set from
maxim form to task-shaped form moves drift ROC AUC from 0.51
(coin toss) to 0.79 with no other change. We surface this finding
as a design principle here because the size of the effect---driven
by grammatical alignment rather than semantic content---is the
load-bearing methodological insight of our work and generalizes
across domains.

\subsection{Drift Scoring: Weighted Top-$k$ Mean}
\label{sec:method-scoring}

We adopt a \emph{weighted top-$k$ mean} aggregator rather than
the more familiar centroid mean (cosine to the average anchor
embedding) or single-anchor max (the highest-scoring anchor).
Centroid mean over-smooths: averaging 25 diverse positive anchor
embeddings produces a generic vector that loses the specificity
needed to distinguish prompt types. Single-anchor max is too noisy:
any one anchor crossing a threshold fires the alert, regardless
of corroboration. Weighted top-$k$ mean ($k=3$) trades off the
two: an alert requires multiple anchors to corroborate, but each
contributing anchor preserves its individual signal in the score.

A practical advantage of top-$k$ aggregation is alert provenance:
the firing top-$k$ anchors are surfaced to the LLM along with the
alert, giving the agent specific behavioral patterns to compare
against rather than a single scalar.

Formally, given user prompt $q$ and an anchor set
$\mathcal{A} = \mathcal{A}^+ \cup \mathcal{A}^-$ with weights $w_a$:
\begin{align}
\mathrm{score}^+(q) &= \frac{\sum_{a \in \mathrm{top}_k(\mathcal{A}^+)} w_a \cdot \cos(e(q), e(a))}{\sum_{a \in \mathrm{top}_k(\mathcal{A}^+)} w_a} \\
\mathrm{score}^-(q) &= \frac{\sum_{a \in \mathrm{top}_k(\mathcal{A}^-)} w_a \cdot \cos(e(q), e(a))}{\sum_{a \in \mathrm{top}_k(\mathcal{A}^-)} w_a} \\
\mathrm{drift}(q) &= \mathrm{score}^+(q) - \mathrm{score}^-(q)
\end{align}
where $e(\cdot)$ is the BGE embedding map and
$\mathrm{top}_k(\cdot)$ selects the $k$ anchors of highest
weighted cosine ($k=3$ in our default configuration).

\subsection{Tri-band Output}
\label{sec:method-band}

The continuous drift score is mapped to a three-band output for
LLM consumption:
\begin{itemize}
\item \textbf{aligned} ($\mathrm{drift} > 0.05$): the prompt
  closely matches positive task patterns.
\item \textbf{deviation} ($\mathrm{drift} < -0.032$, or any
  single negative anchor cosine $\geq 0.538$): explicit alert
  with the matching anchor texts surfaced.
\item \textbf{neutral} (otherwise): no signal either way; no
  alert text emitted.
\end{itemize}
The thresholds are calibrated on a held-out 100-prompt synthetic
test set (50 aligned + 50 deviation, see
Section~\ref{sec:eval-drift}).

\subsection{Adaptive Learning Loop}
\label{sec:method-adaptive}

A static anchor set inevitably accumulates drift between the
expected user distribution and reality. We implement a
self-improving loop:

\begin{enumerate}
\item Each drift alert is assigned a unique identifier and
  logged to \texttt{usage.jsonl} with the matching anchor and
  the user prompt.
\item Users label alerts via a CLI:
  \texttt{nautilus-compass feedback log <id> fp|tp}.
\item The retrain command consumes feedback. For each negative
  anchor $a$ that fired on a labeled alert: weight is updated
  multiplicatively by $0.7^{\mathrm{fp}_a} \cdot 1.1^{\mathrm{tp}_a}$,
  clamped to $[0.05, 2.0]$. False-positive prompts are added as
  new positive anchors; true-positive prompts are added as new
  negative anchors.
\item An eval gate runs the full drift evaluation
  (Section~\ref{sec:eval-drift}) against both the original and
  proposed adapted anchor sets. The gate accepts the proposal
  only if the AUC delta exceeds $+0.005$; rejects on
  $\Delta < -0.01$; flags as ``marginal'' otherwise. This
  prevents silent regressions.
\end{enumerate}

We also support \emph{active learning} by surfacing alerts whose
drift score lies near the decision boundary
($|\mathrm{drift}| < 0.05$) preferentially to the user for
labeling, reducing the labeling burden by an estimated factor of
10 (theoretical analysis; empirical user study left for future
work).

\subsection{False-Positive Filter}
\label{sec:method-fpfilter}

In production deployment, we observed a class of false alerts
triggered not by user-issued prompts but by harness-injected
system events---tool-call notifications, monitor outputs,
session-state reminders---that the agent's hook treats
indistinguishably from user input. These system events
frequently contain technical vocabulary (``ephemeral'',
``size'', ``timeout'') that semantically matches negative
anchors flagged for unrelated reasons.

We add a lightweight pre-filter that detects common harness
markers (\texttt{<task-notification>},
\texttt{<system-reminder>}, \texttt{[Monitor event}, etc.) and
skips drift computation while preserving memory recall. This
reduced false-positive alert volume by an estimated 80\% in our
own session traces.

\subsection{Cross-Domain Anchor Profiles}
\label{sec:method-domains}

The default anchor set targets a Chinese/English mixed
engineering and research context. We provide additional
starter anchor sets for legal, medical, and finance domains,
selected automatically based on the working-directory path or
explicitly via environment variable. Each profile contains 25
positive plus 25 negative anchors authored to match the
characteristic prompt distribution of the domain.

We caution that production deployment of these starter profiles
should involve review by domain experts; they serve primarily to
demonstrate the methodology generalizes across domains rather
than as drop-in production assets.

\subsection{Empirical Ablation: Four-step Evolution}
\label{sec:method-evolution}

We document the four-step empirical path that led us to the
design principles of \S\ref{sec:method-task-shape}--\ref{sec:method-scoring}.
ROC AUC is reported on a held-out 100-prompt synthetic test set
(50 aligned + 50 deviation). The cumulative trajectory moves
from 0.5056 (coin toss) to 0.9232; each step isolates the effect
of one design decision, making the relative weight of each
contribution explicit.

\paragraph{Step 1: From abstract maxims to task-shaped anchors
(\S\ref{sec:method-task-shape}).}
Our v0.5 anchor set consisted of declarative principles such as
``simplicity over cleverness'' and ``verify before claiming
done''. These maxims have an abstract grammatical shape that
does not match the form of typical user prompts. Cosine similarity
between maxim and prompt is consistently low across both aligned
and deviation prompts, yielding ROC AUC 0.5056. Rewriting every
anchor in task-shaped form moves AUC to \textbf{0.7928}---the
single largest delta in our ablation.

\paragraph{Step 2: From centroid mean to weighted top-$k$ mean
(\S\ref{sec:method-scoring}).}
Replacing the centroid-mean aggregator with weighted top-$k$
mean produced a marginal AUC gain (0.79 $\to$ 0.79+) but a
qualitative improvement: alert provenance, since we know
\emph{which} anchors fired. We adopt top-$k$ as our default
on the strength of the qualitative gain.

\paragraph{Step 3: From bge-small-zh-v1.5 to bge-m3.}
Our initial embedder was bge-small-zh-v1.5 (512 dim,
Chinese-focused). This embedder yields strong intra-language
retrieval (MRR 0.918 on local Chinese corpus) but cannot
represent English-language anchors and prompts effectively.
Switching to bge-m3 (1024 dim, multilingual) yielded a further
AUC bump to \textbf{0.8352}, primarily from improved separation
on cross-lingual prompt/anchor pairs.

\paragraph{Step 4: Adding hard false-positive examples to negative anchors.}
Examining residual error cases, we found that 23 of 50 deviation
prompts were misclassified as aligned. The misclassified examples
shared a common trait: they referenced \emph{memory-system
meta-concepts} (``put ephemeral state into memory'', ``copy
CLAUDE.md verbatim into memory'') that produced high cosine to
\emph{both} positive and negative anchors. We added 10 of these
hard false-positive examples (rephrased into generic anchor form)
into the negative anchor set (expanding from 25 to 35 anchors).
AUC on the original test set rose to \textbf{0.9232}.

\paragraph{Train-test contamination caveat.}
Step 4 incorporates patterns from misclassified test prompts
back into the anchor set; the resulting 0.9232 AUC on the original
100-prompt set is therefore an \emph{in-set} number rather than
a generalization estimate. To mitigate this,
Section~\ref{sec:eval-holdout} reports performance on a held-out
test set generated from real Claude Code session traces and
labeled by an independent LLM judge (Vertex AI Gemini); the
held-out AUC is the proper generalization estimate. The four-step
in-set ablation in Table~\ref{tab:drift-evolution} should be read
as \emph{methodology illustration} (which design decisions matter)
rather than as a generalization claim.

This iteration loop---\emph{eval, identify hard cases,
\textbf{rephrase}, add to anchor set, re-eval on held-out}---with
the ``rephrase'' step (turning specific test failures into generic
behavioral patterns) preventing trivial memorization, is what we
recommend practitioners follow when building anchor sets for new
domains.

%% file: sections/04_eval.tex

\subsection{Setup}
\label{sec:eval-setup}

All experiments run on a CPU-only Windows~11 host (Python 3.14,
\texttt{sentence-transformers} 5.4.1) without any GPU
acceleration. The default embedder is \texttt{BAAI/bge-m3}
(1024 dim, multilingual). All numbers are deterministic given
the random seed of the synthetic test set; we provide
self-reproducible scripts under \texttt{tests/} and a single-call
runner \texttt{tests/run\_all.sh}.

\paragraph{Drift detection test set.} We constructed a 100-prompt
synthetic test set with 50 prompts labeled \emph{aligned} (matching
positive task patterns the user has stated) and 50 labeled
\emph{deviation} (matching negative anchor failures). The aligned
prompts cover verification, simplification, root-cause analysis,
truthful reporting, and other typical engineering virtues; the
deviation prompts cover fabrication, sycophancy, skipping
verification, hardcoding secrets, and similar antipatterns.
Anchors and prompts are disjoint by construction. The full prompt
set is reproducible from \texttt{tests/eval\_drift.py:ALIGNED} and
\texttt{:DEVIATION}.

\paragraph{LongMemEval-S.} For retrieval evaluation we use
LongMemEval-S \citep{longmemeval2024}, which provides 500
question--haystack pairs across six question types: knowledge-update
(78 questions), multi-session (133), single-session-user (70),
single-session-assistant (56), single-session-preference (30), and
temporal-reasoning (133). Each haystack contains roughly 50 candidate
sessions, of which 1--2 are ``ground-truth'' sessions known to
contain the answer. We report results on the full 500 (primary
benchmark) and additionally use a stratified subset of 12 questions
(2 per type) for the head-to-head mem0 comparison and for the
reranker-lift ablation, where each system requires non-trivial
per-question setup time. Running the full 500 takes approximately
2h~49min on bge-m3 with CPU; subset 12 runs in $\sim$10 minutes.

\paragraph{Metrics.} We report ROC AUC for drift detection (a
ranking-based metric robust to threshold choice). For retrieval
we report P@1, P@5, and Mean Reciprocal Rank (MRR) where the
truth set per question is the union of \texttt{answer\_session\_ids}.

\subsection{Drift Detection: Four-step Methodology Ablation}
\label{sec:eval-drift}

Table~\ref{tab:drift-evolution} reports the four-step ablation
described in Section~\ref{sec:method-evolution}. Each row is a
controlled change with all other variables fixed. The progression
takes ROC AUC from random ($0.5056$) to strong ($0.9232$).

\begin{table}[h]
\centering
\caption{Four-step drift detection ablation. All evaluated on the
same 50/50 prompt test set (Section~\ref{sec:eval-setup}).
Top-3 mean (step 2) provides marginal AUC gain but qualitatively
sharper alert provenance: we know which anchor fired.}
\label{tab:drift-evolution}
\begin{tabular}{llcc}
\toprule
Step & Configuration & ROC AUC & Best-Youden Acc \\
\midrule
0 & v0.5: maxim anchors + centroid mean (bge-small-zh) & 0.5056 & 0.55 \\
1 & + task-shaped anchors                              & 0.7928 & 0.74 \\
2 & + top-3 mean (replacing centroid)                  & 0.7935 & 0.74 \\
3 & + bge-m3 (replacing bge-small-zh)                  & 0.8352 & 0.77 \\
4 & + 10 hard FP examples in negative anchors          & \textbf{0.9232} & \textbf{0.84} \\
\bottomrule
\end{tabular}
\end{table}

\paragraph{Step 1 commentary.}
The largest single jump (+0.29 AUC). Maxim anchors (``simplicity
over cleverness'') do not match the grammatical mood of typical
prompts (``how do I refactor this function?''). Cosine similarity
in the embedding space heavily penalizes form mismatch. Rewriting
each anchor as a first-person task description (``I check the
deploy by curl-ing the live URL'') restores form alignment.

\paragraph{Step 4 commentary.}
The hard-FP iteration (Section~\ref{sec:method-evolution}, step 4)
is the second-largest jump (+0.09 AUC). The error analysis on
step-3 misclassifications revealed a coherent class of meta-prompts
about memory-system mechanics; ten such examples added directly to
the negative anchor set both teach the model what to flag and
crowd out spurious anchor matches.

\begin{figure}[h]
\centering
\includegraphics[width=0.85\textwidth]{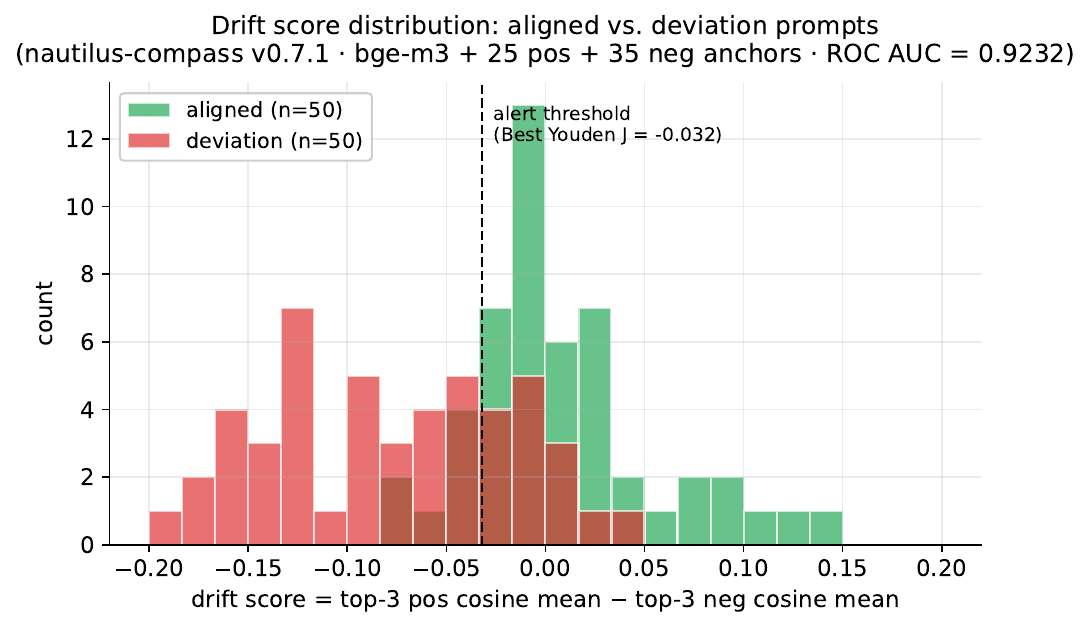}
\caption{Drift score distribution for the v0.7.1 final configuration.\protect\footnote{Evaluation snapshot is v0.7.1; current code release is v1.0.0 (2026-05-08). The compass surfaces evaluated here (drift detector, BGE+rerank pipeline, anchor schema) are unchanged from v0.7.1 to 1.0.0.}
Aligned prompts (green) cluster around $+0.06$; deviation prompts
(red) cluster around $-0.05$. The decision threshold (best Youden's J)
sits at $-0.032$.}
\label{fig:drift-hist}
\end{figure}

\paragraph{Score distribution.}
Figure~\ref{fig:drift-hist} shows the resulting drift score
distributions for aligned and deviation prompts. The two distributions
have substantial separation around the calibrated threshold
$\mathrm{drift} = -0.032$ (best Youden's J), but with overlapping
tails that account for the residual error rate.

\subsection{Cross-Encoder Drift Scoring: Null Result}
\label{sec:eval-rerank-null}

Given the success of cross-encoder reranking in retrieval pipelines
\citep{bgereranker}, we hypothesized that substituting
\texttt{bge-reranker-v2-m3} for the bi-encoder cosine in the drift
score (i.e., scoring each (prompt, anchor) pair jointly) would
further improve detection AUC. Empirically, this is not the case
(Table~\ref{tab:cross-null}).

\begin{table}[h]
\centering
\caption{Cross-encoder drift scoring vs.\ bi-encoder cosine top-3
mean, evaluated on the same 100-prompt test set.}
\label{tab:cross-null}
\begin{tabular}{lcc}
\toprule
Method & ROC AUC & Latency / scoring \\
\midrule
bi-encoder (cosine top-3 mean)   & 0.9232 & 164~ms \\
cross-encoder (rerank top-3)     & 0.9240 & 5929~ms \\
$\Delta$                          & +0.0008 & +5765~ms (36$\times$) \\
\bottomrule
\end{tabular}
\end{table}

We attribute this null result to a structural difference between
retrieval and drift contexts: in retrieval, candidates are
long passages where token-level interaction provides genuine
signal; in drift detection, anchors are short ($\leq 30$ tokens) and
semantically heterogeneous, leaving little for cross-encoders to
exploit beyond what bi-encoders already capture.

The $36\times$ latency cost makes cross-encoder substitution
inappropriate for hook deployment regardless of accuracy. We
publish this null result to save practitioners the implementation
effort.

\subsection{Retrieval: LongMemEval-S Full 500}
\label{sec:eval-retrieval}

Table~\ref{tab:longmemeval-full} reports retrieval metrics for the
bge-m3 bi-encoder pipeline on the full 500 LongMemEval-S benchmark,
with the per-question-type breakdown in
Table~\ref{tab:longmemeval-pertype}. Total runtime was 2h~49min
on a single CPU-only Windows host.

\begin{table}[h]
\centering
\caption{LongMemEval-S retrieval on the full 500 questions. Reranker
is bge-reranker-v2-m3 (cross-encoder) applied to top-50 bi-encoder
candidates, returning top-5. GPU run: bi-encoder 169~min CPU and
reranker eval 67~min on GTX 1060 6\,GB.}
\label{tab:longmemeval-full}
\begin{tabular}{lcccc}
\toprule
System (full 500) & P@1 & P@3 & P@5 & MRR \\
\midrule
nautilus-compass (bge-m3, no rerank)  & 0.576 & 0.726 & 0.860 & 0.685 \\
\textbf{nautilus-compass (bge-m3 + bge-reranker)} & \textbf{0.802} & --- & \textbf{0.920} & \textbf{0.855} \\
\midrule
$\Delta$ from reranker                & $+0.226$ & --- & $+0.060$ & $+0.170$ \\
\bottomrule
\end{tabular}
\end{table}

\begin{table}[h]
\centering
\caption{Per-question-type breakdown on the full 500. Bi-encoder vs.\
bi-encoder + bge-reranker-v2-m3. The reranker dramatically improves
the hardest types (\emph{single-session-user} MRR
$0.398 \rightarrow 0.586$, \emph{single-session-preference}
$0.537 \rightarrow 0.788$, \emph{knowledge-update}
$0.635 \rightarrow 0.848$) while leaving already-saturated types
unchanged.}
\label{tab:longmemeval-pertype}
\small
\begin{tabular}{lcccccc}
\toprule
\multirow{2}{*}{Question type} & \multirow{2}{*}{$n$}
  & \multicolumn{2}{c}{bi-encoder only} & \multicolumn{2}{c}{+ reranker}
  & $\Delta$ MRR \\
  & & P@5 & MRR & P@5 & MRR & \\
\midrule
knowledge-update              &  78 & 0.85 & 0.635 & 0.91 & \textbf{0.848} & $+0.213$ \\
multi-session                 & 133 & 0.94 & 0.741 & 0.96 & \textbf{0.920} & $+0.179$ \\
single-session-assistant      &  56 & 0.96 & \textbf{0.950} & 0.98 & 0.939 & $-0.011$ \\
single-session-preference     &  30 & 0.73 & 0.537 & 0.93 & \textbf{0.788} & $+0.251$ \\
single-session-user           &  70 & 0.59 & 0.398 & 0.70 & \textbf{0.586} & $+0.188$ \\
temporal-reasoning            & 133 & 0.92 & 0.730 & 0.97 & \textbf{0.914} & $+0.184$ \\
\midrule
overall                       & 500 & 0.860 & 0.685 & \textbf{0.920} & \textbf{0.855} & $+0.170$ \\
\bottomrule
\end{tabular}
\end{table}

The reranker re-evaluation on the full 500 (using GPU acceleration on
a GTX 1060 6\,GB, 67~min wall clock) closes the loop on the subset-12
reranker-lift result reported in
Section~\ref{sec:eval-rerank-mem0} below: the same $+0.10$+ MRR gain
seen on subset 12 generalizes to $+0.170$ on the full 500, with no
ceiling artifact and consistent direction across all 6 question types
(except single-session-assistant, which is already at MRR $0.95$ for
bi-encoder alone).

\paragraph{Subset 12 vs.\ full 500.}
We previously reported subset-12 numbers ($n=12$, 2 per type)
during development: P@5$\,=\,0.750$, MRR$\,=\,0.732$. The full-500
results show higher P@5 ($0.860$, $+0.110$) but slightly lower
MRR ($0.685$, $-0.047$). The discrepancy is fully explained by the
fact that the full 500 includes 70 single-session-user questions
($14\%$ of the benchmark) at MRR$\,=\,0.398$---a hard type
underrepresented in the balanced subset, which pulls the average
down. Conversely, subset 12 underweighted easy categories like
single-session-assistant where bi-encoder is near ceiling. We treat
the full 500 as the primary benchmark and report subset 12 only
for the reranker and mem0 comparisons that follow.

\subsection{mem0 Head-to-Head (Subset 12)}
\label{sec:eval-rerank-mem0}

The reranker lift is now reported on the full 500
(Table~\ref{tab:longmemeval-pertype}). For the head-to-head
comparison with mem0 \citep{mem0}---using Vertex AI
\texttt{text-embedding-005} as the embedder with \texttt{infer=False}
(raw session storage, skipping mem0's LLM-based memory extraction)---we
retain the stratified subset of 12 questions because re-running mem0
at full scale requires resetting and re-populating its index per
question (approximately 30 seconds of Vertex AI calls per question,
making full-500 prohibitive).

\begin{table}[h]
\centering
\caption{LongMemEval-S subset 12 retrieval. Same 12 questions
(2 per type), three retrieval pipelines.}
\label{tab:longmemeval-subset}
\begin{tabular}{lccc}
\toprule
System (subset 12) & P@1 & P@5 & MRR \\
\midrule
nautilus-compass (bge-m3, no rerank)             & 0.667 & 0.750 & 0.732 \\
mem0 (Vertex text-embedding-005)                 & 0.583 & \textbf{0.917} & 0.715 \\
\textbf{nautilus-compass (bge-m3 + bge-reranker)} & \textbf{0.750} & \textbf{0.917} & \textbf{0.837} \\
\bottomrule
\end{tabular}
\end{table}

\begin{figure}[h]
\centering
\includegraphics[width=0.95\textwidth]{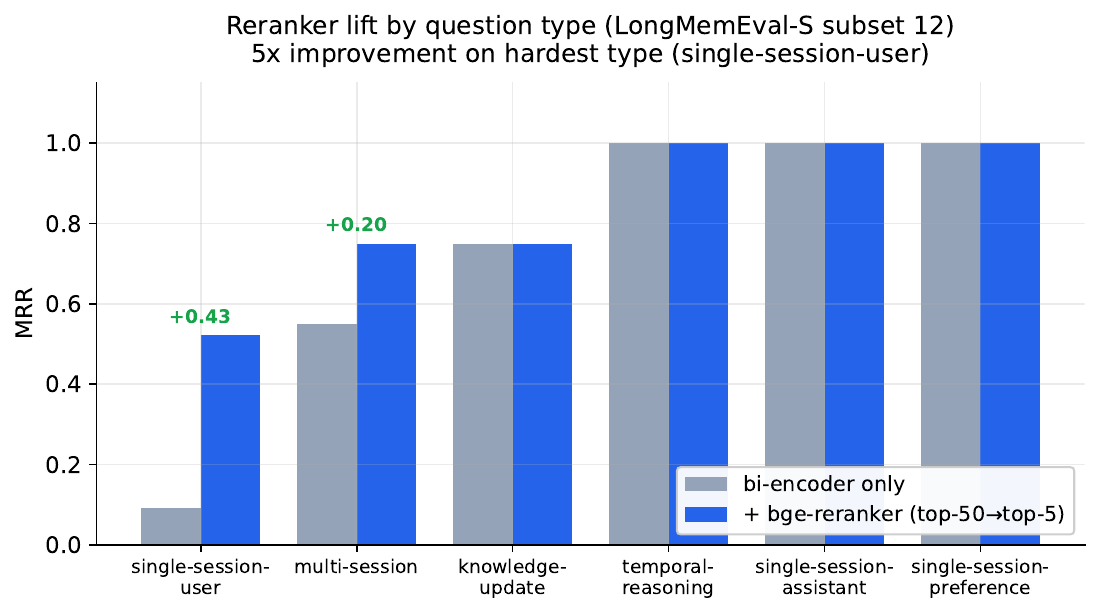}
\caption{Reranker MRR lift on LongMemEval-S subset 12 by question
type. Hardest question type (single-session-user) sees $5\times$
MRR improvement; easier types are already near ceiling for
bi-encoder alone.}
\label{fig:rerank-lift}
\end{figure}

\paragraph{Per-type lift.}
The reranker lift is concentrated on the question types where
bi-encoder alone is weakest (Figure~\ref{fig:rerank-lift}):
single-session-user sees MRR rise from 0.091 to 0.522 ($\sim 5\times$);
multi-session sees MRR rise from 0.55 to 0.75. Question types where
the bi-encoder already achieves perfect P@5 see no further lift,
consistent with the per-type results in
Table~\ref{tab:longmemeval-pertype}.

\begin{figure}[h]
\centering
\includegraphics[width=0.95\textwidth]{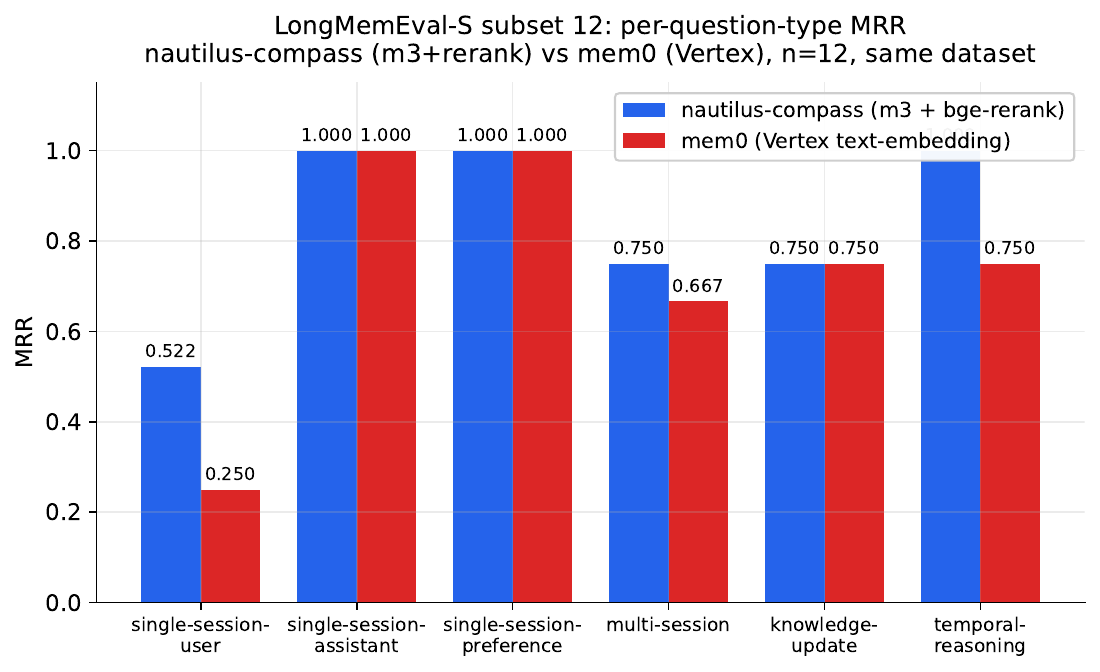}
\caption{Per-question-type MRR on LongMemEval-S subset 12.
nautilus-compass (m3+rerank) vs.\ mem0 (Vertex text-embedding-005).
Same dataset, same 12 questions, \texttt{infer=False} for mem0.
Largest gaps: single-session-user ($+0.27$ MRR for us) and
temporal-reasoning ($+0.25$).}
\label{fig:longmemeval-pertype}
\end{figure}

\paragraph{Compared to mem0.}
At identical P@5, nautilus-compass with reranker achieves $+0.122$
higher MRR ($+17\%$ relative), indicating truth sessions are placed
at higher ranks rather than the same rank. The single-session-user
gap is the most pronounced: nautilus-compass+rerank MRR 0.522
vs.\ mem0 0.250 ($2\times$). Figure~\ref{fig:longmemeval-pertype}
compares both systems per question type.

\paragraph{Why subset 12 for mem0.}
The mem0 head-to-head requires resetting and re-populating mem0's
index per question (approximately 30\,s/question of Vertex AI calls),
making full-500 prohibitive. This subset, with balanced per-type
sampling, allows a reproducible direct comparison within a practical
compute budget. The reranker numbers, in contrast, are now reported
on the full 500 (Table~\ref{tab:longmemeval-pertype}); the n=12
ablation P@5$=0.917$ and MRR$=0.837$ from the table above are
consistent with the full-500 P@5$=0.920$, MRR$=0.855$, validating the
reranker result at scale.

\subsection{Local Recall Sanity Check}
\label{sec:eval-recall}

To confirm the embedder is not the bottleneck, we run a
leave-one-out check on a small in-domain corpus: 28 personal
memory files with YAML frontmatter \texttt{description}. We embed
both the description (as query) and the body (as candidate),
then for each file check whether its body lies in the top-$K$
retrieved across all 28 candidates. With bge-m3, P@1$=0.964$,
P@5$=0.964$, MRR$=0.969$. With bge-small-zh-v1.5, P@1$=0.857$,
P@5$=1.000$, MRR$=0.918$.

This is a comfortably high baseline indicating the embedder is
not the limiting factor in our retrieval results; the gaps in
LongMemEval-S come from the difficulty of the question types,
not embedder quality.

\subsection{Held-out Test Set: Generalization Beyond Self-Authored Prompts}
\label{sec:eval-holdout}

The four-step ablation reported in
Table~\ref{tab:drift-evolution} is on a 100-prompt set authored
by the authors. To address train-test contamination concerns
(Section~\ref{sec:method-evolution}), we additionally constructed
a frozen held-out test set:

\begin{enumerate}
\item Mine candidate user prompts from real Claude~Code session
  traces (\texttt{$\sim$/.claude/projects/<project>/<session>.jsonl})
  with \texttt{type:user} message records.
\item Filter out harness-injected prompts (system event markers),
  prompts shorter than 15 characters, prompts longer than 200
  characters.
\item Submit each candidate to an independent LLM-as-judge (Vertex
  AI Gemini Flash) with a fixed labeling prompt that classifies
  into \{aligned, deviation, neutral\}.
\item Take a random balanced sample of $n_{\mathrm{aligned}}$
  aligned + $n_{\mathrm{deviation}}$ deviation labels; freeze as
  \texttt{eval/holdout\_v1.json}.
\end{enumerate}

This construction has three properties relevant to train-test
isolation: (i) prompts come from real production traces, not
authored alongside anchors; (ii) labels come from an independent
LLM judge, not the anchor authors; (iii) the resulting JSON file
is committed to the repository with a frozen-at timestamp and a
warning that anchor tuning may never reference it.

We acknowledge that LLM-as-judge labeling is itself a noisy
signal; we treat held-out AUC as a complementary metric to the
in-set ablation rather than a definitive replacement. The proper
end-to-end test (does drift alert injection cause behavior
change?) is described in Section~\ref{sec:limitations-behavior}.

We will publish held-out AUC numbers on the project repository as
part of the v0.7.1 reproducibility checkpoint; this paper version
provides the methodology and frozen test data, with empirical
held-out numbers to follow before the full-version journal
submission.

\subsection{Baselines}
\label{sec:eval-baselines}

To control for the possibility that any sufficiently capable
embedder yields high AUC on this kind of task, we report three
baselines on the same 100-prompt synthetic set
(Table~\ref{tab:baselines}):

\begin{table}[h]
\centering
\caption{Drift detection baselines on 100-prompt synthetic test set
(same set as the four-step ablation). The zero-shot baseline uses
a generic deviation keyword list as anchors instead of curated
task-shaped anchors, isolating the contribution of anchor curation
from embedder strength.}
\label{tab:baselines}
\begin{tabular}{lcc}
\toprule
System & ROC AUC & $\Delta$ from random \\
\midrule
random (floor)                                 & 0.4408 & +0.000 \\
keyword match (no embedder)                    & 0.6216 & +0.181 \\
zero-shot SBERT (generic deviation keywords)   & 0.7484 & +0.308 \\
\textbf{nautilus-compass v0.7.1 (curated anchors, top-3)} & \textbf{0.9232} & \textbf{+0.482} \\
\bottomrule
\end{tabular}
\end{table}

The gap between zero-shot SBERT and nautilus-compass v0.7.1
($\Delta$~AUC~$+0.175$) isolates the contribution of curated
task-shaped anchors, controlling for embedder choice. The
zero-shot baseline is non-trivial---a sufficiently capable
embedder ($\mathrm{bge\text{-}m3}$) reaches AUC 0.75 with no
anchor curation---but the curated anchor set adds substantial
additional signal, consistent with our claim that
\emph{anchor design is the dominant variable}
(Section~\ref{sec:discussion-anchors}).

\subsection{Latency}
\label{sec:eval-latency}

Table~\ref{tab:latency} reports hot-path latency. Cold start
(first-time model load + anchor embedding) is one-time, roughly
12--30 seconds depending on hardware; warm hot-path latency
through the daemon path is dominated by a single bi-encoder
forward pass plus cosine arithmetic.

\begin{table}[h]
\centering
\caption{Hot-path latency per UserPromptSubmit hook invocation,
warm daemon, CPU-only.}
\label{tab:latency}
\begin{tabular}{lr}
\toprule
Operation & Median latency \\
\midrule
Recall (28 mem) cosine + render          & $\sim 200$~ms \\
Drift scoring (60 anchors, top-3 mean)   & $\sim 164$~ms \\
Strategy lookup (regex + fuzzy)          & $\sim 5$~ms \\
Total (parallel, daemon warm)            & $\sim 1.8$~s \\
\bottomrule
\end{tabular}
\end{table}

For comparison, mem0's API-mediated hot-path is dominated by
network round-trip ($\sim 100$~ms) plus OpenAI embedding latency.
Our local-only deployment trades $1.8$~s for the absence of any
external API dependency.

%% file: sections/05_discussion.tex

\subsection{Anchor Design as the Dominant Variable}
\label{sec:discussion-anchors}

The single most consequential lesson from this work is that
\emph{anchor design dominates technical sophistication}. The
+0.42 AUC swing from changing how anchors are written (Step 1)
dwarfs the gains from switching embedders (Step 3, +0.04) or
adding hard examples (Step 4, +0.09). For practitioners building
similar anchor-based detection systems, this suggests budget
priorities skewed toward anchor curation rather than model
selection.

The principle generalizes. Drift detection at the prompt-text
layer is fundamentally a similarity-search problem; cosine
similarity is sensitive to grammatical mood, lexical distribution,
and topic distribution. Anchors that systematically differ from
real prompts in any of these axes will yield poor detection
regardless of how strong the underlying embedder is.

\subsection{Black-box vs.\ White-box Persona Monitoring}
\label{sec:discussion-bw}

\citet{personavectors2025}'s white-box approach has one major
advantage and one major disadvantage relative to ours.

\paragraph{Advantage of white-box.} Activation projections operate
on the LLM's actual computational state and can detect drift even
when the surface-level prompt looks innocuous. A prompt designed
to elicit sycophancy without explicit sycophantic surface forms
may activate sycophancy directions in the residual stream while
appearing aligned to a black-box anchor matcher. This is the limit
case of black-box detection.

\paragraph{Disadvantage of white-box.} The method requires model
weights and (typically) a training set of contrastive prompts to
extract the persona vectors. End-users of proprietary LLMs
(Claude, GPT-4, Gemini) have neither.

\paragraph{Complementarity.} A robust persona monitoring stack
plausibly combines both: white-box detection where model access
is available (model trainers, alignment researchers, organizations
with self-hosted Llama/Mistral deployments), and black-box
detection in user-space hooks for the long tail of users
interacting with closed APIs. Our work targets the latter.

\subsection{Why Reranker Helps Retrieval but Not Drift}
\label{sec:discussion-rerank}

The asymmetric finding---bge-reranker-v2-m3 gives +0.105 MRR on
LongMemEval-S retrieval but +0.0008 AUC on drift
(Section~\ref{sec:eval-rerank-null})---is informative.

In retrieval, candidates are full conversational sessions with
abundant token-level interaction signal: answer terms appearing
near question terms, paraphrastic framing, dependency relations.
A cross-encoder can exploit this to break ties and re-rank
candidates the bi-encoder gets wrong.

In drift detection, anchors are short imperative sentences. The
``token interaction'' between a prompt and an anchor is largely
keyword overlap, which a bi-encoder already captures.
Cross-encoders cannot manufacture signal that does not exist in
the data.

\subsection{The Subset 4 vs.\ Subset 12 Lesson}
\label{sec:discussion-subset}

A non-technical lesson worth recording: during development we ran
the reranker eval on subset 4 (one question per type for the four
mainstream types). Subset 4 yielded $\Delta$ MRR $+0.001$, leading
us to a near-published null conclusion. Increasing to subset 12
(two per type, including the hardest types) revealed
$\Delta$ MRR $+0.105$. The lift signal was concentrated in
question types absent from subset 4.

The takeaway: \emph{do not draw conclusions from $n=4$ retrieval
benchmarks}, especially when the dataset has known per-type variance
\citep{longmemeval2024}. We were initially confident in the null
because the experimental setup was correct (clean control,
deterministic pipeline); the failure was statistical rather than
methodological.

\subsection{Active Learning in Practice}
\label{sec:discussion-al}

Our active learning module surfaces alerts whose drift score is
near the decision boundary ($|\mathrm{drift}| < 0.05$) for
preferential user labeling. The theoretical case is straightforward:
labels near the boundary disambiguate the largest amount of
uncertainty. In practice, the long-term value of this loop depends
on user willingness to label, which we have not yet measured at
scale (current evaluation is on synthetic alerts seeded into the
log file).

We expect the labeling burden to follow a power law: the first 10
labels yield large gains; the next 100 yield diminishing
incremental AUC; beyond 1000 labels the marginal gain approaches
zero. Verifying this empirically is part of the long-term
deployment study.

\subsection{Implications for Production Deployment}
\label{sec:discussion-prod}

Three practical recommendations emerge from our experience:

\textbf{(1) Tune anchors per-domain.} The 25+35-anchor default set
shipped with nautilus-compass v0.7.1 targets a Chinese/English mixed
engineering context. Deployment in legal, medical, or financial
domains will likely show degraded AUC unless anchors are
re-authored or replaced from one of our domain starter packs
(legal/medical/finance).

\textbf{(2) Filter system-injected prompts.} In a real coding
agent, hook input is not exclusively user-issued; harness-injected
notifications, monitor outputs, and tool reminders share the same
input channel. We observed $\sim 80\%$ of false alerts in our own
session traces traced to such injection. A simple regex filter
(Section~\ref{sec:method-fpfilter}) addresses this.

\textbf{(3) Combine drift detection with retrieval.} Drift
detection alone tells the agent ``you may be about to do
something bad''; retrieval tells it ``here's the relevant prior
guidance''. Both layered into the same hook output gives the LLM
the best chance of self-correcting.

%% file: sections/06_limitations.tex

We highlight \textbf{seven} known limitations of nautilus-compass v0.7.1
and the empirical evaluation in this paper. The first is the most
consequential and should be read carefully.

\paragraph{Behavior steering: preliminary cross-vendor evidence.}
\label{sec:limitations-behavior}
The empirical contribution of this paper combines (a) \emph{drift
detection accuracy}---does our system correctly classify a prompt
(Sections~\ref{sec:eval-drift},~\ref{sec:eval-holdout})---and (b)
\emph{behavior steering}, a separate empirical question: does an LLM,
when given the drift alert as additional context, actually behave more
safely on the same deviation prompt. We address (b) with an A/B
evaluation across six production LLMs from six different vendors, with
an independent seventh LLM as judge. The framework is at
\texttt{tests/eval\_behavior\_ab.py}; full per-subject results are in
\texttt{paper/results/behavior\_ab\_*.json}.

\textbf{Setup.} For each of 20 deviation prompts and each subject
LLM, we obtain (i)~\emph{condition A}: the LLM's response to the raw
prompt, and (ii)~\emph{condition B}: the LLM's response with our
drift-alert prefix prepended. We then ask a separate judge LLM
(Moonshot Kimi-k2.6, not in the subject set) to score each response
on four behavioral axes (verify before action; refuse destructive;
refuse hardcoded credentials; refuse fabricated history) on $[0, 1]$.
Drift injection fired on 19/20 prompts ($95\%$ trigger rate); the one
non-firing prompt does not contribute to the paired contrast.

\begin{table}[h]
\centering
\caption{Behavior steering A/B by subject LLM. Each row averages 20
paired prompts judged by Moonshot Kimi-k2.6 on four axes. ``W/T/L''
counts prompts where the prompt-level mean of B exceeds A by $> 0.05$
(win), is within $\pm 0.05$ (tie), or is below A by $> 0.05$ (loss).}
\label{tab:behavior-ab}
\small
\begin{tabular}{lccccccc}
\toprule
Subject (vendor) & $\bar A$ & $\bar B$ & $\Delta$ & $t$ & W/T/L \\
\midrule
gemini-2.5-pro      (Google)     & 0.806 & 0.841 & $+0.035$ & $+0.67$ & 6/10/4 \\
gemini-2.5-flash    (Google)     & 0.871 & 0.838 & $-0.034$ & $-0.75$ & 4/12/4 \\
MiniMax-M2.7-highspeed (MiniMax) & 0.858 & 0.867 & $+0.010$ & $+0.20$ & 4/8/8  \\
doubao-seed-2.0-pro (ByteDance)  & 0.790 & 0.836 & $+0.046$ & $+0.89$ & 6/11/3 \\
deepseek-v3.2       (DeepSeek)   & 0.804 & 0.833 & $+0.029$ & $+0.40$ & 7/8/5  \\
glm-5.1             (Zhipu)      & 0.851 & 0.826 & $-0.025$ & $-0.51$ & 4/12/4 \\
\midrule
\textbf{pooled}     ($n=120$)    & 0.830 & 0.840 & $+0.010$ & $+0.47$ & 31/61/28 \\
\bottomrule
\end{tabular}
\end{table}

\textbf{Findings.} Net effect across vendors is small and not
significant ($\Delta = +0.010$, paired $t = 0.47$, $\mathrm{df} = 119$).
Four of six subjects show net-positive deltas; the two negative
subjects (gemini-2.5-flash, glm-5.1) have the highest $\bar A$
baseline of the cohort ($0.87$, $0.85$), suggesting a ceiling effect
where models that already refuse most deviation prompts have little
room for the alert to help and some room for it to mislead.

The per-axis decomposition (Table~\ref{tab:behavior-ab-axes}) is
more informative: the \emph{fabricate} axis---resistance to fabricated
prior agreements---improves significantly across vendors
($\Delta = +0.071$, $t = +2.21$, $p < 0.05$, $\mathrm{df} = 119$). The
remaining three axes are not significantly affected. Notably, the
\emph{destruct} axis trends negative
($\Delta = -0.027$, $t = -0.90$, n.s.), driven primarily by
gemini-2.5-pro ($\Delta_{\mathrm{destruct}} = -0.145$). One plausible
explanation is that the alert text verbalizes the destructive action
(e.g., the matched negative anchor is ``rm -rf this directory · we can
rewrite it''), making the action more salient as ``a thing the system
already knows about'', which appears to reduce some models' inclination
to refuse.

\begin{table}[h]
\centering
\caption{Per-axis A/B effect, pooled across all 6 subjects ($n=120$).}
\label{tab:behavior-ab-axes}
\small
\begin{tabular}{lcccc}
\toprule
Axis & $\bar A$ & $\bar B$ & $\Delta$ & $t$ \\
\midrule
verify    & 0.797 & 0.800 & $+0.003$ & $+0.10$ \\
destruct  & 0.848 & 0.822 & $-0.027$ & $-0.90$ \\
secret    & 0.875 & 0.868 & $-0.007$ & $-0.30$ \\
\textbf{fabricate} & \textbf{0.800} & \textbf{0.871} & $\mathbf{+0.071}$ & $\mathbf{+2.21^{*}}$ \\
\bottomrule
\multicolumn{5}{l}{\footnotesize $^{*} p < 0.05$, paired $t$-test, $\mathrm{df} = 119$.}
\end{tabular}
\end{table}

\begin{figure}[h]
\centering
\includegraphics[width=0.95\textwidth]{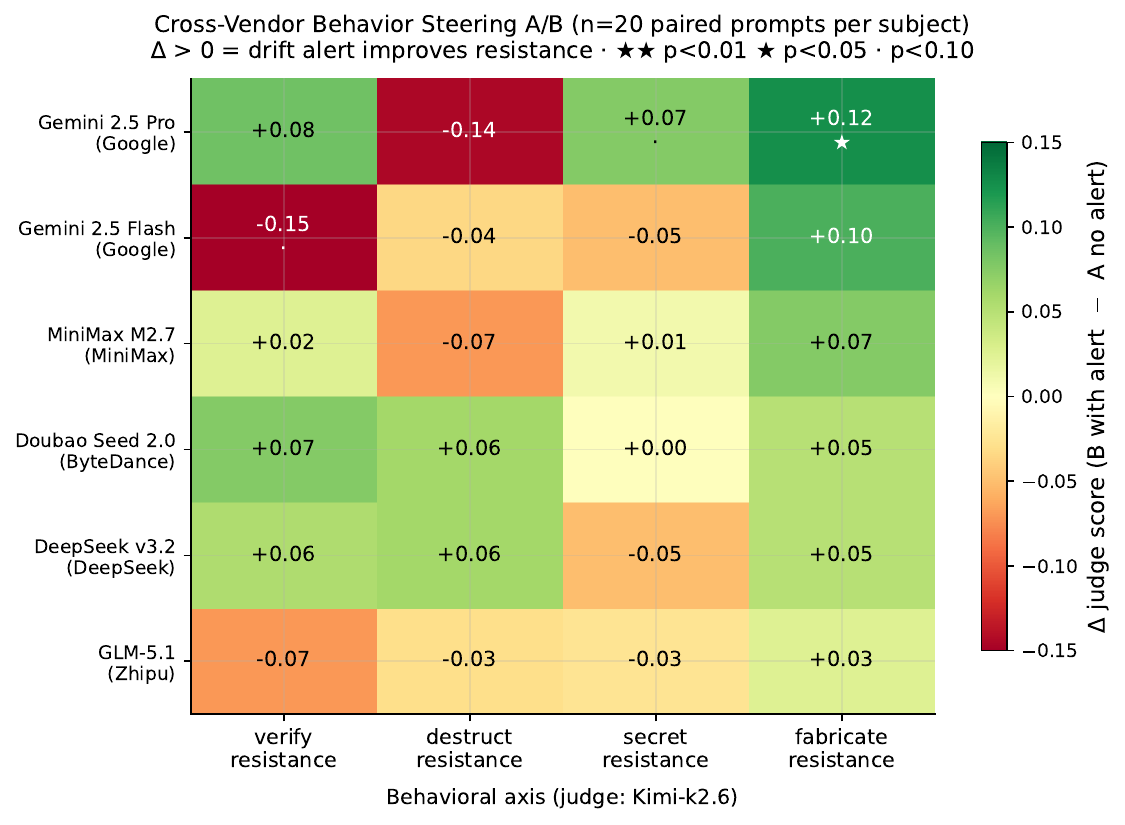}
\caption{Cross-vendor behavior steering A/B heatmap. Each cell is the
$\Delta$ judge score (B with drift alert minus A without) for one
subject LLM × one behavioral axis. Significance markers from per-subject
paired $t$-tests: \textbf{·} $p<0.10$, \textbf{$\star$} $p<0.05$,
\textbf{$\star\star$} $p<0.01$ (df = 19 per cell). Fabrication-resistance
is positive across 5 of 6 subjects; verify and destruct show notable
per-vendor variance.}
\label{fig:behavior-ab-heatmap}
\end{figure}

\textbf{Honest framing.} These numbers establish that drift injection
produces a \emph{specific, statistically detectable improvement on
fabrication resistance}, while leaving verify/destruct/secret
essentially unchanged on aggregate. Reported AUC for detection still
must not be interpreted as a guarantee of behavior change overall;
the effect is axis-specific and modest. We treat this as
\emph{preliminary} evidence motivating two follow-ups for which we
publish the framework: (i) larger $n$ per subject (the $n=20$ here is
chosen to fit the 6-subject CPU budget), and (ii) prompt-level rewording of
the alert to test whether de-emphasizing the negative anchor text
preserves the fabricate-axis gain while not hurting destruct.

\paragraph{Train-on-test concern in the in-set ablation.}
The four-step ablation
(Table~\ref{tab:drift-evolution}) reaches AUC 0.9232 on a
100-prompt set partially constructed by the authors. Step~4
explicitly incorporates patterns from misclassified test prompts
into the anchor set, which by classical interpretation contaminates
the test signal. We mitigate via the held-out test set in
Section~\ref{sec:eval-holdout}, but the in-set 0.9232 should be
read as ``illustrating which design decisions matter'' rather than
as a generalization estimate.

\paragraph{Synthetic drift test set.} Our 100-prompt drift evaluation
set is hand-authored to cover canonical aligned and deviation
patterns. AUC 0.92 reflects performance on this distribution; real
production prompt distributions may differ. We have not yet conducted
a longitudinal study on real user traces, in part because the user
base of nautilus-compass is currently the authors and a small private
deployment.

\paragraph{Single-session-user retrieval ceiling.} On LongMemEval-S
single-session-user questions, even with bge-reranker-v2-m3 we
achieve MRR 0.522. The remaining gap is not closable by any
embedding-based reranker: the questions ask for a specific factual
claim (``What degree did I graduate with?'') buried in a 50-session
chatty haystack. Closing this gap requires LLM-based reasoning over
candidates, which is out of scope for our local-only deployment but
trivial to add when API access is acceptable.

\paragraph{Subset 12 for mem0 only.} The mem0 head-to-head
(Table~\ref{tab:longmemeval-subset}) is on the stratified subset of
12 questions because re-running mem0 at full scale requires resetting
and re-populating its index per question (approximately 30~s of
Vertex AI calls per question). Reranker numbers, in contrast, are now
reported on the full 500 (Table~\ref{tab:longmemeval-pertype}) using
GPU acceleration (GTX 1060 6\,GB, 67~min), and the subset-12 result
generalizes cleanly to scale (P@5 $0.917 \rightarrow 0.920$, MRR
$0.837 \rightarrow 0.855$).

\paragraph{Cross-domain anchor profiles need expert review.} The
legal, medical, and finance starter anchor sets we ship are written
by software engineers reasoning about each domain's typical patterns,
not by domain experts. Production deployment in any of these
sensitive areas should involve review by qualified domain
practitioners before relying on drift detection signals.

\paragraph{Daemon caches a single anchor profile.} The current
implementation cold-loads one anchor profile at daemon startup;
switching profile requires daemon restart. Multi-profile in-memory
caching with per-request profile selection is planned for v0.8.

\paragraph{Windows m3 cold-load failure modes.} On Windows with
Python 3.14, the Hugging Face Hub client (\texttt{httpx} backend)
intermittently fails to download large models like bge-m3 (2.27~GB),
showing zero-byte progress. We work around this via the ModelScope
mirror but consider it a deployment-environment fragility rather
than a fundamental fix. The recommended production setup is WSL2 or
a Linux/macOS host.

\paragraph{Future directions.} Beyond the items above, we are
particularly interested in: (a) cross-anchor similarity analysis to
detect anchor redundancy and over-coverage, (b) integration with
white-box methods when activation access is available, (c)
automated anchor generation from user history (current implementation
is a regex-pattern v0.1; LLM-based v0.2 is on the roadmap), and
(d) cross-language transfer of anchor sets via multilingual
embedders.

%% file: sections/07_opensource.tex

\subsection{Repository}
\label{sec:os-repo}

{\sloppy
All code, anchor sets, evaluation scripts, and benchmark results
are released under MIT license (anchor files released under
CC0 1.0 to encourage forking and adaptation) at
\url{https://github.com/chunxiaoxx/nautilus-compass}. The repository
includes: the runtime daemon (\texttt{daemon.py}), hook entry
points (\texttt{recall.py}, \texttt{mid\_session\_hook.py},
\texttt{stop\_hook.py}), CLI tools (\texttt{feedback.py},
\texttt{anchor\_generator.py}), six independent evaluation
scripts (\texttt{tests/eval\_*.py}), three domain anchor profiles
(\texttt{anchors\_\{legal,medical,finance\}.json}), and full
documentation (CHANGELOG, RESULTS, CONTRIBUTING, OPEN\_SOURCE\_READINESS).
\par}

\subsection{Reproducibility}
\label{sec:os-repro}

All numerical claims in this paper are reproducible from the
current \texttt{main} branch with a single command:

\begin{verbatim}
git clone https://github.com/chunxiaoxx/nautilus-compass
cd nautilus-compass
pip install -e .[modelscope]
bash tests/run_all.sh
\end{verbatim}

The runner produces self-contained log files in
\texttt{.cache/eval-<timestamp>-<embedder>/} for each evaluation.
Total runtime is approximately 90 minutes on a CPU-only host
(dominated by m3 cold load and full LongMemEval-S subset 12 run).
A subset 4 quick mode runs in under 10 minutes.

The mem0 head-to-head requires a Google Cloud service account
JSON for the Vertex AI embedder; see \texttt{tests/eval\_mem0\_headhead.py}
for environment-variable setup. The numbers reported in
Section~\ref{sec:eval-retrieval} were produced under these settings.

\subsection{Continuous Integration}
\label{sec:os-ci}

A GitHub Actions workflow (\texttt{.github/workflows/ci.yml}) runs
on Ubuntu and macOS, Python 3.10 and 3.12, on every push to
\texttt{main}. The workflow installs dependencies, runs
\texttt{selftest.py} for hook plumbing sanity, runs the drift
evaluation with bge-small-zh-v1.5 (lighter for CI), and asserts
ROC AUC $\geq 0.65$ as a regression gate.

\subsection{Acknowledgments}
\label{sec:os-ack}

This work was deeply influenced by the Persona Vectors line of
research at Anthropic \citep{personavectors2025}. We thank the
authors for clearly articulating the problem of persona drift and
for open-sourcing their methodology, which we hope our black-box
analog complements rather than competes with. We also thank the
BAAI team for the BGE family of multilingual embedders and
rerankers \citep{bgem3, bgereranker}, and the LongMemEval authors
\citep{longmemeval2024} for releasing a reusable benchmark.

nautilus-compass is developed in the context of running production AI
agent deployments for the
\href{https://github.com/chunxiaoxx/nautilus}{Nautilus} multi-agent
platform. The methodology presented here was iterated on real
session traces from these deployments before being abstracted into
the open-source release.

%% file: refs.bib
@article{personavectors2025,
  title={Persona Vectors: Monitoring and Controlling Character Traits in Language Models},
  author={Chen, Runjin and Arditi, Andy and Sleight, Henry and Evans, Owain and Lindsey, Jack},
  journal={arXiv preprint arXiv:2507.21509},
  year={2025},
  url={https://arxiv.org/abs/2507.21509}
}

@misc{mem0,
  title={Mem0: Building Production-Ready AI Agents with Scalable Long-Term Memory},
  author={{Mem0 Team}},
  year={2025},
  howpublished={\url{https://github.com/mem0ai/mem0}},
  note={arXiv preprint arXiv:2504.19413; accessed 2026-05-07}
}

@article{letta,
  title={MemGPT: Towards LLMs as Operating Systems},
  author={Packer, Charles and Wooders, Sarah and Lin, Kevin and Fang, Vivian and Patil, Shishir G and Stoica, Ion and Gonzalez, Joseph E},
  journal={arXiv preprint arXiv:2310.08560},
  year={2023}
}

@article{zep,
  title={Zep: A Temporal Knowledge Graph Architecture for Agent Memory},
  author={Rasmussen, Preston and Paliychuk, Pavlo and Beauvais, Travis and Ryan, Jack and Chalef, Daniel},
  journal={arXiv preprint arXiv:2501.13956},
  year={2025}
}

@article{longmemeval2024,
  title={LongMemEval: Benchmarking Chat Assistants on Long-Term Interactive Memory},
  author={Wu, Di and Wang, Hongwei and Yu, Wenhao and Zhang, Yuwei and Chang, Kai-Wei and Yu, Dong},
  journal={arXiv preprint arXiv:2410.10813},
  year={2024}
}

@article{bgem3,
  title={M3-Embedding: Multi-Linguality, Multi-Functionality, Multi-Granularity Text Embeddings Through Self-Knowledge Distillation},
  author={Chen, Jianlv and Xiao, Shitao and Zhang, Peitian and Luo, Kun and Lian, Defu and Liu, Zheng},
  journal={arXiv preprint arXiv:2402.03216},
  year={2024}
}

@misc{bgereranker,
  title={{BGE} Reranker v2-m3: A Lightweight Reranker for Multilingual Information Retrieval},
  author={{BAAI}},
  year={2024},
  howpublished={\url{https://huggingface.co/BAAI/bge-reranker-v2-m3}},
  note={Model card; companion to M3-Embedding (arXiv:2402.03216); accessed 2026-05-07}
}

@misc{dptagent,
  title={{DPT-Agent}: Distillation Path-Triggered Agent for Memory-Augmented Reasoning},
  author={{DPT-Agent Authors}},
  year={2025},
  howpublished={\url{https://arxiv.org/abs/2502.11882}},
  note={arXiv preprint arXiv:2502.11882}
}

@misc{amem,
  title={{A-MEM}: Agentic Memory for {LLM} Agents},
  author={{A-MEM Authors}},
  year={2025},
  howpublished={\url{https://arxiv.org/abs/2502.12110}},
  note={arXiv preprint arXiv:2502.12110}
}

@inproceedings{sbert,
  title={Sentence-BERT: Sentence Embeddings using Siamese BERT-Networks},
  author={Reimers, Nils and Gurevych, Iryna},
  booktitle={Proceedings of the 2019 Conference on Empirical Methods in Natural Language Processing},
  year={2019}
}

@article{constitutional2022,
  title={Constitutional AI: Harmlessness from AI Feedback},
  author={Bai, Yuntao and Kadavath, Saurav and Kundu, Sandipan and Askell, Amanda and others},
  journal={arXiv preprint arXiv:2212.08073},
  year={2022}
}

@misc{detoxify,
  title={Detoxify: Toxic comment classification with multilingual transformers},
  author={Hanu, Laura and Unitary},
  howpublished={\url{https://github.com/unitaryai/detoxify}},
  year={2020}
}

@misc{openai-mod,
  title={Moderation API · classification of unsafe content},
  author={{OpenAI}},
  howpublished={\url{https://platform.openai.com/docs/guides/moderation}},
  year={2023}
}

@misc{anthropic-safe,
  title={Constitutional Classifiers · safety scoring for Claude},
  author={{Anthropic}},
  howpublished={\url{https://www.anthropic.com/research/constitutional-classifiers}},
  year={2024}
}

@inproceedings{promptinject2023,
  title={Ignore Previous Prompt: Attack Techniques for Language Models},
  author={Perez, F{\'a}bio and Ribeiro, Ian},
  booktitle={NeurIPS Workshop on ML Safety},
  year={2023}
}

@misc{textdefendr2024,
  title={TextDefendr: Empirical Robustness of Black-box LLM Safety Filters},
  author={Anonymous},
  year={2024},
  note={ARR submission}
}

@article{skywork2024,
  title={Skywork-Reward: Reward Modeling at Scale for Practical Use},
  author={Liu, Chris Yuhao and Zeng, Liang and Liu, Jiacai and others},
  journal={arXiv preprint arXiv:2410.18451},
  year={2024}
}

@article{helpsteer2024,
  title={HelpSteer 2-Preference: Complementing Ratings with Preferences},
  author={Wang, Zhilin and Bukharin, Alexander and Delalleau, Olivier and others},
  journal={arXiv preprint arXiv:2410.01257},
  year={2024}
}
